\documentclass{article}
\usepackage{fullpage}
\usepackage{alltt}
\usepackage{array}
\usepackage{moreverb}
\usepackage[dvips]{epsfig}

\title{Early Experience with ASDL in lcc}

\author{David R. Hanson\\
Microsoft Research\\
1 Microsoft Way, Redmond, WA 98052\\
{\normalsize\texttt{drh@microsoft.com}}}

\date{MSR-TR-98-50\\September, 1998}
\newcommand{\singlespace}{\renewcommand{\baselinestretch}{1.0}\normalsize}

\begin{document}
\maketitle
\begin{abstract}
The Abstract Syntax Description Language (ASDL) is a language for
specifying the tree data structures often found in compiler intermediate
representations. The ASDL generator reads an ASDL specification and
generates code to construct, read, and write instances of the trees
specified. Using ASDL permits a compiler to be
decomposed into semi-independent components that communicate by reading
and writing trees. Each component can be
written in a different language, because the ASDL generator can emit
code in several languages, and the files written by ASDL-generated
code are machine- and language-independent.
ASDL is part of the National Compiler
Infrastructure project, which seeks to reduce
dramatically the overhead of computer systems research by
making it much easier to build high-quality compilers.
This paper describes dividing lcc, a widely used retargetable C
compiler, into two components that communicate via trees defined in
ASDL. As the first use of ASDL in a `real' compiler, this
experience reveals much about the effort required to
retrofit an existing compiler to use ASDL, the overheads involved,
and the strengths and weaknesses of ASDL itself and,
secondarily, of lcc.

\end{abstract}
\bibliographystyle{abbrv}

\setcounter{secnumdepth}{0}
\section{Introduction}

High-quality compilers for a range of modern languages are
essential for conducting experimental research in computer
architecture, programming languages, and programming environments.
For example, compilers are
required to run benchmarks for evaluating new ideas in architecture
and code optimization. And compilers for new languages need
optimizers, code generators, and runtime systems for existing
platforms. 

Building compilers is often a bottleneck in
these kinds of research projects because compiler construction is a
labor-intensive activity. Often, nearly complete compilers must be
constructed even if the essential components are relatively small
parts of the whole. To evaluate a new architecture, for instance,
requires a code generator for that architecture and perhaps an
architecture-dependent optimizer. Writing toy compilers for toy
languages is insufficient: The research community demands measurements using
established benchmarks, like the SPEC
benchmarks~\cite{spec95}, which are written in real programming
languages.

The National Compiler Infrastructure (NCI) project seeks to reduce
dramatically the effort needed to perform realistic experiments by
making it much easier to build high-quality compilers.
The goal is to make it possible to build complete compilation systems from
pieces, replacing or modifying only those components that are relevant to
the client researchers. For example, researchers studying global
optimization algorithms for C++ would replace or add only their optimizers
and would use existing C++ front ends and code generators.

The NCI can pay both economic and intellectual
dividends. It should reduce significantly the costs of doing computer
systems research, in terms of time, barrier to entry, and direct
monetary outlay. It should also encourage more researchers to attack
computer systems problems and thus increase the rate at which new
research results appear.

The NCI includes the Stanford Intermediate
Format~\cite{wilson:etal:94} and the emerging Zephyr
program-generation tools, which includes the Abstract Syntax
Description Language (ASDL)~\cite{wang:etal:97}.
ASDL describes the abstract syntax of compiler intermediate
representations and other tree-like data structures. The
ASDL generator, asdlGen, converts ASDL specifications
into appropriate data-structure definitions, constructors, and
functions to read and write these data structures to files in a
variety of programming languages.

This paper describes how ASDL is used with lcc~\cite{fraser:hanson:95}, a
well-documented, small, production-quality compiler for ISO Standard
C~\cite{harbison:steele:95}. This experience is valuable for two
reasons. First, lcc is perhaps the simplest C compiler available and
thus provides a `basis' test case for ASDL and other NCI tools. If ASDL
can't handle lcc's intermediate representation, it's unlikely to work
in more ambitious compilers or in compilers for higher-level languages.
Second, lcc wasn't designed to be decomposed into reusable program
components, so doing so suggests how difficult it is to retrofit ASDL into
existing compilers.

\section{ASDL}

ASDL is a small, domain-specific language for describing tree data
structures~\cite{wang:etal:97}. ASDL specifications are concise and independent of
any particular programming language. The ASDL generator, asdlGen,
accepts an ASDL specification and emits code that defines a concrete
representation for the data structures described in the specification,
along with code that constructs, reads, and writes instances of those
data structures. Currently, asdlGen can emit data-structure
implementations in C, C++, Java, ML, and Haskell. ASDL specifications
tend to be much smaller than the corresponding, language-specific
data structure and function definitions.

\begin{figure}
\begin{center}
\epsfig{file=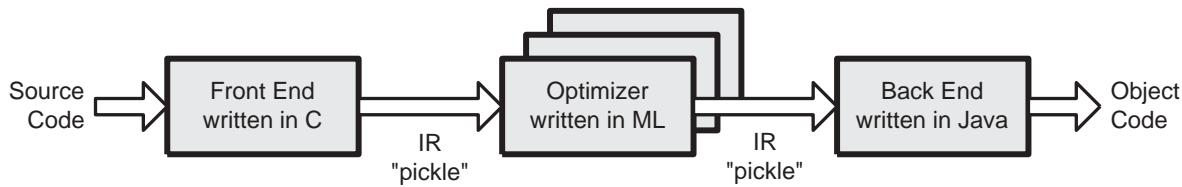}
\end{center}
\caption{Sample compiler organization using ASDL.}
\label{fig:asdl}
\end{figure}

Compiler writers can use ASDL to partition a
compiler into several independent programs as depicted in
Fig.~\ref{fig:asdl}. A front end reads source code and builds an
intermediate representation using the data structure constructors
generated by asdlGen. It writes these data structures to a file---a
`pickle'---using the I/O functions generated by asdlGen.
Subsequent phases read and write pickles as necessary, perhaps
modifying them in the process. For example, optimizers would read a
pickle, improve the code therein, and write a new pickle.

The binary pickle format is independent of both language and host
platform. Thus, as suggested in Fig.~\ref{fig:asdl}, compiler
phases can be written in whatever ASDL-supported language best suits the task at
hand. If ASDL becomes widely used, researchers can tap into a
complete compiler by adding new phases or replacing just the phases
of interest.

ASDL is not a universal intermediate representation~\cite{steel61}, because
it supports any IR that can be described by trees. Likewise, ASDL is
not a universal distribution format~\cite{andf-spec}, because it does
not mandate specific formats, capabilities, or platforms. ASDL and asdlGen are to compilers
what interface definition languages (IDLs)~\cite{snodgrass89} and
stub generators are to distributed systems. Typical IDLs describe the interfaces between
program components running in different address spaces, and stub
generators generate implementations of these functions that use
remote procedure calls to communicate between clients and servers.
ASDL describes the data-structure interfaces between compiler phases,
and asdlGen generates functions to communicate between these phases.

ASDL is such simple language that examples suffice to explain nearly all
of its features. The following ASDL specification describes an IR
for a language of arithmetic expressions, assignment statements, and
print statements.
\begin{quote}
\begin{verbatimtab}\singlespace
module IR {
stm	= SEQ(stm, stm)
	| ASGN(identifier, exp)
	| PRINT(exp*)

exp	= OP(binop, exp, exp)
	| ID(identifier)
	| ICON(int)
	| RCON(real)

real	= (int, int)

binop	= ADD | SUB | MUL | DIV
}
\end{verbatimtab}
\end{quote}
This specification defines four types: \texttt{stm}, \texttt{exp},
\texttt{real}, and \texttt{binop}. The first three productions define
the sum type \texttt{stm}, which has three
\emph{constructors}. A \texttt{stm} is a \texttt{SEQ} tree with two \texttt{stm}
children, an \texttt{ASGN} tree with two children of types
\texttt{identifier} and \texttt{exp}, or a \texttt{PRINT} tree with
one child of type list of \texttt{exp}. \texttt{identifier} is a
built-in type, and the `\texttt{*}' following a type specifies a
list of that type.

Similarly, \texttt{exp} is a sum type with four constructors that
describe trees for binary operators, identifiers, and integer and
real constants. \texttt{int} is another built-in type, but there is
no built-in type for reals. So, the \texttt{real} type is a
product type whose instances represent real numbers as two
integers. Finally, \texttt{binop} is a simple sum type that
defines constructors for each of the possible binary operators.

All four types are wrapped in a module named \texttt{IR}; this
name is used to provide a disambiguating prefix for the names in the
generated implementations.

It is easy to confuse ASDL specifications with grammars for
programming languages. This ASDL specification describes the
\emph{abstract syntax} of the intermediate representation for
programs written in some unspecified concrete syntax.

Given an ASDL specification, asdlGen emits an interface and an
implementation in the programming language specified. The interface
defines the language-specific representation for the types and
declares functions for constructing instances of those types and
for reading and writing them. The functions themselves appear in the
implementation. For languages that do not separate interfaces and
implementations, like Java, asdlGen emits a single implementation.

In C, for example, given the ASDL specification above in the file
\texttt{IR.asdl}, asdlGen writes the interface to \texttt{IR.h} and
the implementation to \texttt{IR.c}. Figure~\ref{fig:IR.h} shows the
snippets from \texttt{IR.h} that define the representation for
\texttt{stm} and the associated constructors, readers, and writers.
\begin{figure}
\begin{center}
\begin{minipage}{0.75\linewidth}
\begin{verbatimtab}\singlespace
...
struct IR_stm_s {
    enum {IR_SEQ_enum, IR_ASGN_enum, IR_PRINT_enum} kind;
    union  {
        struct IR_SEQ_s { IR_stm_ty stm1; IR_stm_ty stm2;} IR_SEQ;
        struct IR_ASGN_s {
            identifier_ty identifier1;
            IR_exp_ty exp1;
        } IR_ASGN;
        struct IR_PRINT_s { list_ty exp_list1;} IR_PRINT;
    } v;
};
...
IR_stm_ty IR_SEQ(IR_stm_ty stm1, IR_stm_ty stm2);
IR_stm_ty IR_ASGN(identifier_ty identifier1, IR_exp_ty exp1);
IR_stm_ty IR_PRINT(list_ty exp_list1);
...
extern IR_stm_ty IR_read_stm(instream_ty s_);
extern void IR_write_stm(IR_stm_ty x_, outstream_ty s_);
\end{verbatimtab}
\end{minipage}
\end{center}
\caption{Generated C interface for the example ASDL specification.}
\label{fig:IR.h}
\end{figure}
ASDL uses compact and efficient representations whenever possible.
A sum type is represented by a union. There is one field for
each constructor and a corresponding function for building instances
of that constructor, as shown for \texttt{stm} in
Fig.~\ref{fig:IR.h}. ASDL represents simple sum types
with integers or their language-specific equivalent. In C, for example,
\texttt{binop} is represented by just an enumeration type.
The implementation, \texttt{IR.c}, contains the definitions for the
functions declared in \texttt{IR.h}.

ASDL comes with libraries of basic types and functions for
each programming language it supports. These libraries provide
support for lists and for the built-in types, such as
\texttt{identifier}, in languages that do not support them directly.
In C, lists are represented by an implementation
of variable-length sequences~\cite[Ch.~11]{hanson:97}, and identifiers are
represented by a C implementation of atoms~\cite[Ch.~3]{hanson:97}.

As Fig.~\ref{fig:IR.h} reveals, asdlGen generates field names and
parameter names as necessary to complete the data structure and
function definitions. Compiler writers can specify these names in the
ASDL specification. For example, if \texttt{stm} is defined as
\begin{quote}
\begin{verbatimtab}\singlespace
stm	= SEQ(stm first, stm rest)
	| ASGN(identifier id, exp e)
	| PRINT(exp* elist)
\end{verbatimtab}
\end{quote}
\texttt{first},
\texttt{rest},
\texttt{id},
\texttt{e}, and
\texttt{elist} will be used for the corresponding field and parameter
names in Fig.~\ref{fig:IR.h}.

Sum types can also have \emph{attributes}, which are fields that are
common to all constructors. For example,
\begin{quote}
\begin{verbatimtab}\singlespace
stm	= SEQ(stm first, stm rest)
	| ASGN(identifier id, exp e)
	| PRINT(exp* elist)
	  attributes(int lineno)
\end{verbatimtab}
\end{quote}
attaches a line number attribute to each constructor. Attributes are
usually factored into a common prefix for the type, e.g., the C type
for \texttt{stm} from Fig.~\ref{fig:IR.h} becomes
\begin{quote}
\begin{verbatimtab}\singlespace
struct IR_stm_s {
    int_ty lineno;
    enum {IR_SEQ_enum, IR_ASGN_enum, IR_PRINT_enum} kind;
    union  {
        struct IR_SEQ_s { IR_stm_ty first; IR_stm_ty rest;} IR_SEQ;
        struct IR_ASGN_s { identifier_ty id; IR_exp_ty e;} IR_ASGN;
        struct IR_PRINT_s { list_ty elist;} IR_PRINT;
    } v;
};
\end{verbatimtab}
\end{quote}

\section{The lcc Code-Generation Interface}

lcc is a retargetable compiler for ISO Standard C. It is
distributed with back ends for the SPARC, MIPS, X86, and ALPHA for a
variety of platforms. Others have written back ends for additional
platforms, and lcc is used by other compiler researchers; for
example, a modified, older release of lcc is used as the C compiler in the
SUIF project.

Communication between lcc's target-independent front end and its
target-dependent back ends is specified by a small code-generation
interface. This interface consists of a few shared
data structures, a 33-operator tree IR that represents executable code, and
18 functions that manipulate trees and the shared data structures.

The shared data structures include tree nodes, symbol-table entries,
and types. The 33 tree IR operators are listed in
Fig.~\ref{fig:operators}. Each of these generic operators can be
specialized by appending an operand type suffix and a size in bytes.
The 6 type suffixes are:
\begin{quote}\singlespace
\begin{tabular}{>{\tt}ll}
F & float \\
I & integer \\
U & unsigned \\
P & pointer \\
B & `block' (aggregate) \\
V & void
\end{tabular}
\end{quote}
There can be up to 9 sizes. For example, \texttt{ADDF4} denotes a
4-byte floating addition, and \texttt{CVII2} denotes a conversion
from an integer to a 2-byte integer.
While it looks like $33\times 6\times 9 = 1782$ specific operators, not all combinations are 
meaningful, and the number of sizes on most targets is limited. On 32-bit targets, 
there are 130 type- and size-specific operators.
Conversions on 32-bit targets, for instance, convert only between 4
and 4- or 8-byte floats, or widen or narrow between 3 sizes of integers.
Some operators have only one or a few valid suffixes; for
instance, the address operators \texttt{ADDRL}, \texttt{ADDRF}, and
\texttt{ADDRG} can have only the `\texttt{P}' type suffix and whatever size is
the size of a pointer on the target. Back end authors need
accommodate only those type- and size-specific operators that are meaningful on
their targets.

\begin{figure}[b]
\begin{center}\tt\singlespace
\begin{tabular}{lllllllll}
CNST	& ARG	& ASGN	& INDIR	& CVF	& CVI	& CVP	& CVU \\
NEG	& CALL	& RET	& ADDRG	& ADDRF	& ADDRL	& ADD	& SUB \\
LSH	& MOD	& RSH	& BAND	& BCOM	& BOR	& BXOR	& DIV \\
MUL	& EQ	& GE	& GT	& LE	& LT	& NE	& JUMP	& LABEL
\end{tabular}
\end{center}
\caption{lcc tree IR generic operators.}
\label{fig:operators}
\end{figure}

Incidentally, the lcc 3.x interface~\cite{fraser:hanson:95} supported
only three sizes of integers, two sizes 
of floats, and insisted that pointers fit in unsigned integers. These assumptions 
simplified the compiler and were suitable for 32-bit architectures,
but not for 64-bit architectures. The main difference between the 3.x interface and the 4.x
interface described here are the operator size suffixes.

Figure~\ref{fig:functions} summarizes the purpose of the 18
code-generation functions. On most targets, implementations of many
of these functions are very short, perhaps only a few calls to
\texttt{printf}, because they simply emit assembly language.
Most of the work goes into \texttt{gen}, \texttt{emit}, and
\texttt{function}, which collaborate to generate and emit code for a function.
While not required by the interface, all of lcc's
distributed back ends use a variant of the IBURG code-generator
generator~\cite{fraser:hanson:proebsting:92a} to specify instruction
selection. The resulting code generators emit optimal local code.
Instruction selection specifications and target-dependent functions
run about 700 lines per target. There are about another 900 lines of
code that are shared between all targets and include functions for
register allocation, etc.

\begin{figure}
\begin{center}\singlespace
\begin{tabular}{>{\tt}ll}
void progbeg(int, char *[])	& initialize the back end \\
void progend(void)		& finalize the back end \\
void defsymbol(Symbol)		& initialize a symbol-table entry \\
void export(Symbol)		& export a symbol \\
void import(Symbol)		& import a symbol \\
void global(Symbol)		& define a global \\
void local(Symbol)		& define a local \\
void address(Symbol, Symbol, long)	& define an address relative to a symbol \\
void blockbeg(Env *)		& open a block-level scope \\
void blockend(Env *)		& close a block-level scope \\
\multicolumn{2}{l}{\tt void function(Symbol, Symbol [], Symbol [], int)} \\
					& define a function body \\
void gen(Node)			& generate code \\
void emit(Node)			& emit code \\
void defconst(int, int, Value)	& initialize a arithmetic constant \\
void defaddress(Symbol)		& initialize an address constant \\
void defstring(int, char *)	& initialize a string constant \\
void space(int)			& define an uninitialized block \\
void segment(int)		& switch logical segments
\end{tabular}
\end{center}
\caption{lcc code-generation functions.}
\label{fig:functions}
\end{figure}
\begin{figure}[!t]
\begin{center}
\epsfig{file=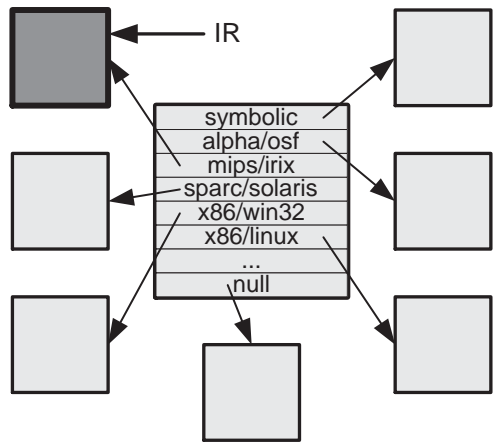}
\end{center}
\caption{Specifying a target, e.g., \texttt{mips/irix}, selects an interface record.}
\label{fig:IR}
\end{figure}


lcc's packaging is somewhat novel: Pointers to the code-generation
functions and some target-specific parameters are packaged 
in the following `interface record:'
\begin{quote}\singlespace
\begin{alltt}
struct interface \{
        Metrics charmetric, shortmetric, intmetric, longmetric, \textrm{\ldots};
        unsigned little_endian:1, mulops_calls:1, wants_callb:1, \textrm{\ldots};
        void (*address)(Symbol, Symbol, long);
        void (*blockbeg)(Env *);
        void (*blockend)(Env *);
        \textrm{\ldots}
\};
\end{alltt}
\end{quote}
The \texttt{Metrics} values give the sizes and alignments of the basic
data types, and the 1-bit flags identify other target-dependent
features, like endianness.
There is one interface record for each distinct target, but 
different records can share functions.
lcc is a small compiler, so all of the back ends are combined into a
single executable program, which makes lcc a cross compiler.
As depicted in Fig.~\ref{fig:IR}, a command-line option selects the desired target, e.g.,
\begin{quote}\singlespace
\begin{verbatim}
lcc -Wf-target=mips/irix -S wf1.c
\end{verbatim}
\end{quote}
causes lcc to compile \texttt{wf1.c} and leave the generated MIPS
assembly code in \texttt{wf1.s}.
The \texttt{-Wf-target} option points \texttt{IR} to the appropriate
interface record, and the front end makes indirect calls to the code-generation functions, e.g., 
\begin{quote}\singlespace
\begin{verbatim}
(*IR->defsymbol)(p);
\end{verbatim}
\end{quote}
The default target is the host, so this option is required only for cross compilation.

\begin{figure}
\begin{center}
\epsfig{file=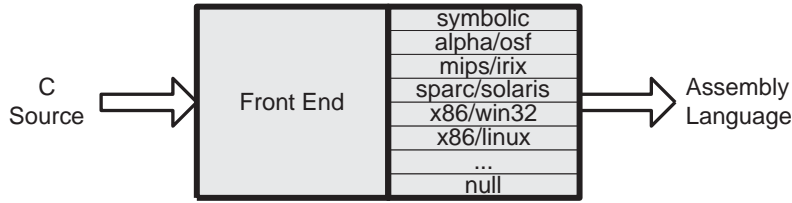}
\end{center}
\caption{lcc's monolithic design: One front end, numerous back ends.}
\label{fig:design}
\end{figure}

\section{Dividing lcc}

lcc is---by design---a monolithic compiler: The front end and the
back ends are combined into a single address space, so the front and back
ends communicate by function calls that exchange pointers to shared data
structures, as Fig.~\ref{fig:design} illustrates.
Also, back ends can make upcalls to functions provided by the front end.
There are about a half dozen such functions, e.g., data-structure 
constructors, a memory allocator, type predicates, and
so on.

lcc is small, at least in comparison with other compilers, because
it omits some components, most notably a global optimizer. One way to
add more functionality is to split lcc into a separate front end and
one or more separate back ends so that an optimizer can be
run between these programs. This design would also make it
easier to use lcc in research projects.

Splitting lcc into two separate programs requires either massive
revisions or some way to read and write the data and actions
represented by the existing code-generation interface.
ASDL facilitates the second alternative: It helps divide lcc into separate programs with 
\emph{no} change to the code-generation interface. So, the existing back ends can be 
used unmodified.

\begin{figure}
\begin{center}
\epsfig{file=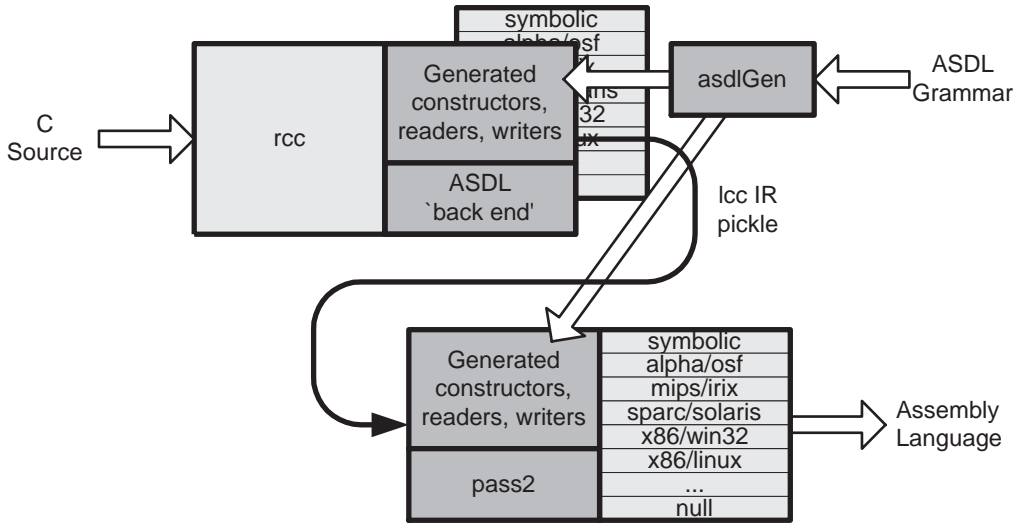}
\end{center}
\caption{lcc's revised design: Separate front end and back ends.}
\label{fig:pass2}
\end{figure}

Figure~\ref{fig:pass2} depicts this revised design.
The front end, rcc, emits a pickle that encodes all the data structures and 
the function calls made when compiling a C source file.
The new program, dubbed `pass2,' reads a pickle, recreates the internal 
data structures, and makes the function calls encoded in the pickle. The 
generated assembly language is often byte-for-byte identical to the code emitted by 
the monolithic compiler. Differences occur only when the back end
calls the label generator, which causes the revised design to number
labels differently.

asdlGen reads the lcc-specific ASDL grammar described in the next section and emits C code for
the data-structure constructors, readers, and writers, which is
included in both rcc and pass2.
Otherwise, the revised front end, rcc, is nearly identical to the
original front end. The ASDL emission 
is accomplished by an ASDL back end, which `spoofs' the back end by 
overwriting the target-specific code-generation function pointers with pointers to 
its own functions.
The back ends are actually linked into both rcc and pass2, because the interface 
records carry important machine parameterizations required by both programs. 
This packaging is not essential; rcc could link in the interface records without the 
functions.

The revised compiler is used much like the original, except that ASDL
output must be specified to rcc, and
pass2 must be run to emit the generated code, e.g.,
\begin{quote}\singlespace
\begin{alltt}
lcc -Wf-target=mips/irix -Wf-asdl -S wf1.c
mv wf1.s wf1.pickle
\textrm{\ldots}
pass2 wf1.pickle >wf1.s
\end{alltt}
\end{quote}
The first command runs rcc and leaves the pickle in \texttt{wf1.s},
which the second command renames. The third command generates the
MIPS assembly code in \texttt{wf1.s}, perhaps after it has been
optimized or otherwise processed by an intervening step.

\subsection{The ASDL Grammar}

The ASDL grammar for lcc is small---only about 70 lines.
It specifies data structures that are target-dependent, which means, 
for example, that it is impossible to specify, say, the
\texttt{mips/irix} target to rcc and the \texttt{x86/win32}
target to pass2. Indeed, pass2 does not accept a target option,
because the target specification is embedded in the pickle.

Figures~\ref{fig:grammar}, \ref{fig:type}, \ref{fig:node},
and~\ref{fig:interface} show the complete lcc ASDL grammar.
The grammar specifies more than just what is in the code-generation 
interface, because pass2 must recreate the compilation environment built by rcc, 
the front end. There are two important ramifications of this requirement.
First, the pickles include complete type information, for example, everything 
about structures and unions, etc. The code-generation interface includes only the 6 basic 
types. This information is defined by the \texttt{type} sum type.
Second, lcc's code-generation data structures and related internal structures are 
graphs, not trees. Thus, items with multiple references are identified by integers, 
and the references to them replaced by these integers. Fields named
\texttt{uid} in the grammar identify these integers.
Dealing with graphs is ASDL's major shortcoming.

\begin{figure}[!b]
\begin{center}
\begin{minipage}{0.75\linewidth}
\begin{alltt}\singlespace
module rcc \{

program = (int nuids,int nlabels,item* items,interface* interfaces,
                int argc,string *argv)

item    = Symbol(symbol symbol)
        | Type(type type)
          attributes(int uid)

symbol  = (identifier id,int type,int scope,int sclass,
                int ref,int flags)

\textrm{See Fig.\ \ref{fig:type}\ldots}

\textrm{See Fig.\ \ref{fig:interface}\ldots}

\textrm{See Fig.\ \ref{fig:node}\ldots}

\}
\end{alltt}
\end{minipage}
\end{center}
\caption{The ASDL grammar for lcc's code-generation interface.}
\label{fig:grammar}
\end{figure}

The product type \texttt{program} is the first ASDL type in
Fig.~\ref{fig:grammar}, and an instance of this type represents a C
compilation unit. That is, rcc `compiles' a C source file into a
\texttt{program} and writes it to a pickle, which pass2 reads and
traverses to generate code. The \texttt{program} type
carries counts of the number of unique integers, uids for short, and
the number of generated labels, a sequence of \texttt{item} types, a sequence of
\texttt{interface} types, the command-line argument count, and the arguments
themselves. The type \texttt{string} is an ASDL built-in type.
The sum type \texttt{item}
carries a uid (as an attribute) and either the associated symbol or
type. The \texttt{item} sequence in a \texttt{program} associates uids with \texttt{symbol}s and
\texttt{type}s, as described below.

Symbol-table entries are an example of a multiply referenced data
structure. Symbol-table entries are 
represented by the product type \texttt{symbol} (see
Fig.~\ref{fig:grammar}), which is a straightforward rendition of lcc's
internal symbol-table entry. It carries the symbol's name, type, scope, 
storage class, how often it's referenced, and some flags.
For example, the C declaration
\begin{quote}\singlespace
\begin{alltt}
struct elem \{ int count; struct elem *left, *right; char *word; \} *root; 
\end{alltt}
\end{quote}
declares \texttt{root} to be a pointer to a struct elem. The
corresponding \texttt{symbol} is
\begin{quote}\singlespace
\begin{alltt}
(id = root, type = \textsl{10}, scope = LOCAL, sclass = AUTO,
        ref = 120000, flags = addressed) 
\end{alltt}
\end{quote}
where, for clarity, symbolic values appear for the \texttt{scope},
\texttt{sclass}, and \texttt{flags} fields. The \texttt{id} field is
an instance of the built-in ASDL type \texttt{identifier}, which are atoms.
The \texttt{type} field---`\texttt{\slshape 10}' in this example---is
a uid that identifies a \texttt{type} value
defined somewhere else in the \texttt{item} sequence. Here and below,
uids are shown in a slanted typewriter font.

\subsection{Types}

Types are represented by instances of the sum type \texttt{type}
defined in Fig.~\ref{fig:type} and are essentially abstract syntax
trees of the C type constructors.
For example, \texttt{INT} is a basic type; \texttt{POINTER} represents a pointer type and its 
integer field is the uid of the referent type; and \texttt{STRUCT} represents a structure type with a tag 
and an ordered set of fields.
The fields are represented by a sequence of \texttt{field} product types, one for each 
field, giving the field name, its type, offset, and location information for bit fields.
Other types are similarly represented.
Every type has attributes that give its size and alignment constraint
in bytes.

A snippet of the \texttt{item} sequence for the \texttt{type}
representing the C type struct elem defined previously
helps clarify the definition of uids and their use:
\begin{quote}\singlespace
\begin{alltt}
\textsl{11}: STRUCT( size = 16, align = 4, fields = [
          id   type  offset  bitsize  lsb
        (count, \textsl{12},     0,      0,     0),
        (left,  \textsl{10},     4,      0,     0),
        (right, \textsl{10},     8,      0,     0),
        (word,  \textsl{13},    12,      0,     0)] )
\textsl{12}: INT(    size =  4, align = 4)
\textsl{10}: POINTER(size =  4, align = 4, type = \textsl{11})
\textsl{13}: POINTER(size =  4, align = 4, type =  \textsl{8})
 \textsl{8}: INT(    size =  1, align = 1)
\end{alltt}
\end{quote}
Again, the italicized numbers are uids. The uids on the left are the
\texttt{uid} attributes in the \texttt{item} type, and each of these define
a uid and its associated \texttt{type}. The occurrences of uids in a
\texttt{type} field are references to types. Type 11 is the \texttt{type}
value for `structtype;' its fields give the size of instances of
this struct (16 bytes), their alignment (on 4-byte boundaries), and
their fields. Each of the \texttt{field} values in the sequence
include a uid for the \texttt{type} of that field.
Type 10 is the C type `struct~elem~*.'
Notice the two kinds of \texttt{INT}s: Type 12 is a 4-byte integer, which is the C type `int,' 
and type 8 is a 1-byte integer, which is type C type `signed char.'
Thus, type 13 is the C type `char~*.'

\begin{figure}
\begin{center}
\begin{minipage}{0.75\linewidth}
\begin{verbatimtab}\singlespace
field	= (identifier id,int type,int offset,int bitsize,int lsb)

enum	= (identifier id,int value)

type	= INT
	| UNSIGNED
	| FLOAT
	| VOID
	| POINTER(int type)
	| ENUM(identifier tag,enum* ids)
	| STRUCT(identifier tag,field* fields)
	| UNION(identifier tag,field* fields)
	| ARRAY(int type)
	| FUNCTION(int type,int* formals)
	| CONST(int type)
	| VOLATILE(int type)
	  attributes(int size,int align)
\end{verbatimtab}
\end{minipage}
\end{center}
\caption{ASDL grammar for C types.}
\label{fig:type}
\end{figure}

\subsection{IR Trees}

IR trees are represented by a nearly isomorphic set of trees defined
by the ASDL sum type \texttt{node}, defined in Fig.~\ref{fig:node}.
Some generic operators are represented by corresponding constructors,
e.g., \texttt{CNST} and \texttt{ADDRL}. Others are represented by
constructors for a class of generic operators in which the specific
operator is provided as a parameter: \texttt{CVT} nodes represent the
conversion operators (\texttt{CVF}, \texttt{CVI}, \texttt{CVP}, and
\texttt{CVU}), \texttt{Unary} and \texttt{Binary} nodes represent the
unary (\texttt{INDIR}, \texttt{RET}, \texttt{JUMP}, \texttt{NEG},
\texttt{BCOM}) and binary operators (\texttt{ADD}, \texttt{SUB},
\texttt{DIV}, \texttt{MUL}, \texttt{MOD}, \texttt{BOR},
\texttt{BAND}, \texttt{BXOR}, \texttt{RSH}, \texttt{LSH}), and
\texttt{Compare} nodes represent the comparisons (\texttt{EQ}, \texttt{NE},
\texttt{GT}, \texttt{GE}, \texttt{LE}, \texttt{LT}).

\texttt{LABEL}
nodes are label definitions, and \texttt{BRANCH} nodes are
unconditional jumps. \texttt{Compare}, \texttt{LABEL}, and
\texttt{BRANCH} nodes use label numbers instead of symbol-table
entries for labels; pass2 recreates the symbol-table entries as it
reconstructs the IR. \texttt{CSE} nodes identify common
subexpressions and associate a symbol-table entry for a temporary
(\texttt{uid}) with a node that computes the subexpression
(\texttt{node}). Every node includes suffix and size attributes,
which correspond to the type and size suffixes in the type- and
size-specific IR operators.

It is important to realize that \texttt{node}s are not lcc IR trees---they represent IR 
trees. In pass2, nodes provide the data necessary to recreate the lcc IR trees, 
which are passed to the back ends. This `duplication of effort' is an
onerous side effect of retrofitting an existing compiler with ASDL
and is discussed in more detail below.

\begin{figure}[!t]\begin{center}
\begin{minipage}{0.75\linewidth}
\begin{verbatimtab}\singlespace
node	= CNST(int value)
	| CNSTF(real value)
	| ARG(node left,int len,int align)
	| ASGN(node left,node right,int len,int align)
	| CVT(int op,node left,int fromsize)
	| CALL(node left,int type)
	| CALLB(node left,node right,int type)
	| RET
	| ADDRG(int uid)
	| ADDRL(int uid)
	| ADDRF(int uid)
	| Unary(int op,node left)
	| Binary(int op,node left,node right)
	| Compare(int op,node left,node right,int label)
	| LABEL(int label)
	| BRANCH(int label)
	| CSE(int uid,node node)
	  attributes(int suffix,int size)
\end{verbatimtab}
\end{minipage}
\end{center}
\caption{ASDL grammar for IR trees.}
\label{fig:node}
\end{figure}
\begin{figure}
\begin{center}
\epsfig{file=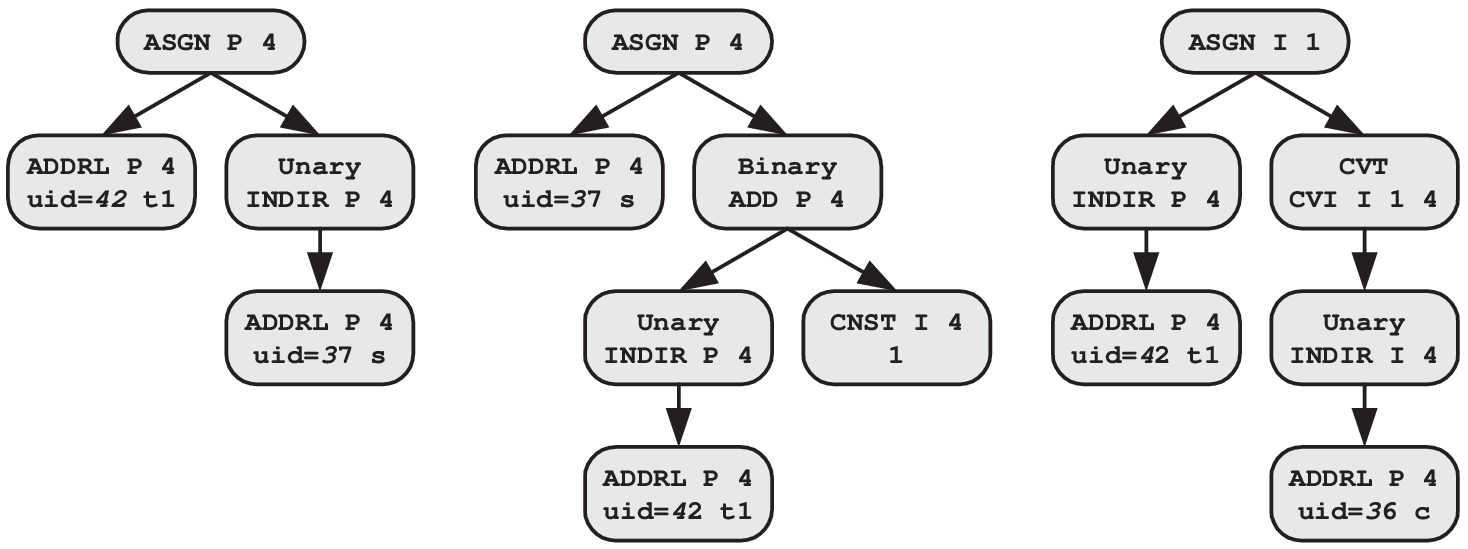}
\end{center}
\caption{ASDL representation for \texttt{t1 = s; s = t1 + 1; *t1 = c}.}
\label{fig:IRtrees}
\end{figure}


lcc compiles the C code
\begin{quote}\singlespace
\begin{alltt}
char *s; int c; *s++ = c;
\end{alltt}
\end{quote}
into the equivalent of
\begin{quote}\singlespace
\begin{alltt}
t1 = s; s = t1 + 1; *t1 = c;
\end{alltt}
\end{quote}
where \texttt{t1} is a compiler-generated temporary.
Figure~\ref{fig:IRtrees} shows the ASDL \texttt{node}s for these three
statements on a 32-bit target.
The notation \texttt{ASGN}~\texttt{P}~\texttt{4} gives the
constructor, the suffix attribute as one of the types listed above, and the size attribute.
Notice the constructor for the indirection in the leftmost tree;
it's a \texttt{Unary} node with three values: 
operator \texttt{INDIR}, suffix \texttt{P}, and size \texttt{4}. The
node for \texttt{Binary} is similar.
Leaves, like \texttt{ADDRL}, include the uid of the appropriate
symbol. For clarity, Fig.~\ref{fig:IRtrees} shows 
the names, too, e.g., \texttt{t1}, but the names are not in the node.

Nodes (and all ASDL-defined data structures) are written to pickles
in a compact, prefix, binary representation in which integers can
take as little as one byte. For example, the leftmost
tree in Fig.~\ref{fig:IRtrees} takes 15 bytes:
\begin{quote}\singlespace
\begin{alltt}
ASGN P 4 ADDRL P 4 42 Unary INDIR P 4 ADDRL P 4 37
\end{alltt}
\end{quote}

\subsection{Interface Calls}

The ASDL types described above encode the data structures in lcc's
code-generation interface. The ASDL sum type \texttt{interface},
defined in Fig.~\ref{fig:interface}, encodes the calls made from the
front end to the back end. Compare this type definition with the
interface calls listed in Fig.~\ref{fig:functions}. The only significant change is that 
symbol-table pointers have been replaced by the corresponding uids or 
sequences of uids. \texttt{progbeg} and \texttt{progend} have been
omitted because pass2 can simply make these calls; pass2 also
supplies the actual arguments for \texttt{blockbeg} and
\texttt{blockend}, so these arguments are not included in
\texttt{interface}. Calls to \texttt{defconst} with real
values are represented by a separate constructor, \texttt{Defconstf},
which confines the use of real values. ASDL has no built-in support
for reals, so they are represented with integers for their
most significant and least significant bits.

\texttt{Address} and \texttt{Local} constructors associate uids with
`relative' symbols and locals and parameters. A relative symbol is
one that is defined by a constant offset from another symbol, e.g.,
\texttt{a[i+10]} would elicit a definition of a symbol to represent
\texttt{\&a[10]}.
Instances of \texttt{Forest} carry the \texttt{node}s that represent the
executable code in each function. These appear in the
\texttt{interface} list in the
\texttt{codelist} field of \texttt{function}s. There is one
\texttt{function} for each C function in the input.

\begin{figure}
\begin{center}
\begin{minipage}{0.75\linewidth}
\begin{verbatimtab}\singlespace
real		= (int msb,int lsb)

interface	= Export(int p)
		| Import(int p)
		| Global(int p,int seg)
		| Local(int uid,symbol p)
		| Address(int uid,symbol q,int p,int n)
		| Segment(int seg)
		| Defaddress(int p)
		| Deflabel(int label)
		| Defconst(int suffix,int size,int value)
		| Defconstf(int size,real value)
		| Defstring(string s)
		| Space(int n)
		| Function(int f,int* caller,int* callee,
			int ncalls,interface* codelist)
		| Blockbeg
		| Blockend
		| Forest(node* nodes)
\end{verbatimtab}
\end{minipage}
\end{center}
\caption{ASDL grammar for code-generation interface calls.}
\label{fig:interface}
\end{figure}

Here's an example: The small function
\begin{quote}\singlespace
\begin{verbatimtab}
err(s) char *s; {
	printf("? %s\n", s);
	exit(1);
}
\end{verbatimtab}
\end{quote}
yields the following sequence of \texttt{interface}s:
\begin{quote}\singlespace
\begin{alltt}
Local(uid = \textsl{27}, symbol = (id = s, \textrm{\ldots}))
Local(uid = \textsl{28}, symbol = (id = s, \textrm{\ldots}))
Function(f = \textsl{22} err, caller = [ \textsl{27} ], callee = [ \textsl{28} ],
        ncalls = 2, codelist = [
                Blockbeg,
                Forest(nodes = \textrm{\ldots}),
                Forest(nodes = \textrm{\ldots}),
                Blockend,
                Forest(nodes = \textrm{\ldots})])
\textrm{\ldots}
Global(p = \texttt{24}, seg = LIT)
Defstring("? %s\char92n")
\end{alltt}
\end{quote}
The \texttt{Local}s above define two views of the formal parameter
\texttt{s}, one as seen by callers of \texttt{err} and one as seen by
the callee itself. These have the same name---\texttt{s}---but are
different symbols as indicated by their different uids.
Often, these symbols are identical, but there are
important cases when they are different, as detailed below.
The first two occurrences of \texttt{Forest} carry the \texttt{node}s
for the two function calls. The third \texttt{Forest} holds a single \texttt{LABEL}
node that marks the location of the function epilogue.
\texttt{Global} and \texttt{Defstring} collaborate to initialize
a compiler-generated static variable for the format string shown.

\section{Measurements}

While retrofitting lcc to use ASDL changed lcc's structure
dramatically, this process did not add much code.
The ASDL grammar described in the previous section is
about 70 lines, the ASDL back end is 409 lines of C, and pass2 is 681
lines of C.

The 70-line ASDL grammar generates about 2183 lines of C declarations and
function definitions. This code is approximately what would be
required if the ASDL-generated constructors, readers, writers were written
by hand. The savings would increase if the ASDL grammar were used to
generate code in other languages. For example, if a Java optimizer
were written, it would use the 3332 lines of Java generated from
the same ASDL grammar.

On Windows NT, the size of the monolithic compiler executable is
380~KB (produced by the Microsoft Visual C/C++ 5.0 compiler with --O1
optimization). The revised rcc with the ASDL back end and the
generated constructors, readers, and writers is 437~KB, and pass2 is
395~KB. Rcc includes the back ends for all of lcc's targets, because
that packaging is the simplest one. The code for these back ends and
the symbol-table emission code could be omitted saving about 199~KB.

Table~\ref{table:results} summarizes the compilation times and the file
sizes for the monolithic and divided variants of lcc when compiling
its own non-trivial modules. The times are given in centiseconds and are
for the compilation phase only; that is, the timings do not include
preprocessor and assembler times. All timings were taken on a lightly
loaded 200MHz Gateway PC with 128~MB of RAM and SCSI disks running
Windows NT 4.0. The compiler variants were compiled by the Microsoft
Visual C/C++ 5.0 compiler with --O1 optimization.

\begin{table}[!t]\singlespace
\caption{Compilation times and output file sizes.}
\label{table:results}
\begin{center}\small
\newcommand{\z}{\hspace{-0.5em}}\newcommand{\Z}{\hspace{0.5em}}
\begin{tabular}{>{\hspace{0.5em}}c>{\hspace{0.5em}}c>{\hspace{0.5em}}cccc>{\ttfamily}l}
\noalign{\hrule height 1pt\kern 2pt}
\texttt{lcc -S}
	& \texttt{lcc -S -asdl} 
		& \texttt{pass2} 
			& Pickle 
				& Object file 
					& Object file 
						& File \\
100ths sec.
	& 100ths sec.
		& 100ths sec.
			& size in KB 
				& size in KB 
					& w/symbols in KB \\ \hline
\Z8	& 20	& \Z6	& 22.2	& \Z1.7	& \Z3.4	& alloc.c \\
\z115	& \z211	& 98	& \z242.1	& 91.5	& \z124.4	& alpha.c \\
15	& 33	& 13	& 37.7	& 11.8	& 24.1	& bytecode.c \\
33	& 61	& 31	& 81.7	& 26.0	& 45.4	& dag.c \\
45	& 67	& 35	& 90.5	& 33.9	& 48.5	& dagcheck.c \\
42	& 77	& 44	& \z100.8	& 36.9	& 61.4	& decl.c \\
35	& 65	& 34	& 82.6	& 20.6	& 36.7	& enode.c \\
26	& 25	& \Z8	& 27.4	& \Z5.3	& 12.2	& error.c \\
48	& 65	& 34	& 81.2	& 25.3	& 44.2	& expr.c \\
44	& 67	& 33	& 77.0	& 23.7	& 47.0	& gen.c \\
24	& 40	& 20	& 40.5	& \Z8.4	& 20.0	& init.c \\
16	& 26	& \Z8	& 26.5	& \Z4.9	& 11.2	& input.c \\
28	& 43	& 21	& 58.9	& 20.2	& 30.9	& lex.c \\
22	& 32	& 12	& 35.7	& \Z8.8	& 19.8	& main.c \\
\z119	& \z161	& 87	& \z206.2	& 75.6	& \z106.4	& mips.c \\
35	& 32	& 10	& 33.1	& \Z7.5	& 14.7	& output.c \\
18	& 28	& 11	& 34.7	& \Z7.1	& 18.5	& prof.c \\
19	& 33	& 13	& 34.4	& \Z5.3	& 11.6	& profio.c \\
55	& 69	& 30	& 82.6	& 26.8	& 69.1	& rcc.c \\
45	& 65	& 34	& 92.2	& 22.4	& 37.3	& simp.c \\
\z147	& \z221	& \z157	& \z264.9	& 94.1	& \z129.9	& sparc.c \\
44	& 35	& 13	& 40.1	& 11.0	& 22.9	& stab.c \\
40	& 52	& 25	& 63.5	& 20.1	& 39.6	& stmt.c \\
16	& 28	& \Z8	& 25.5	& \Z2.7	& \Z4.8	& string.c \\
25	& 31	& 11	& 35.6	& 11.8	& 26.2	& sym.c \\
45	& 51	& 18	& 55.1	& 22.7	& 43.0	& symbolic.c \\
17	& 32	& 12	& 32.4	& \Z8.0	& 18.5	& trace.c \\
44	& 32	& 12	& 35.2	& \Z7.6	& 15.1	& tree.c \\
40	& 59	& 31	& 79.7	& 28.9	& 49.5	& types.c \\
\z185	& \z249	& \z177	& \z298.0	& 97.3	& \z133.9	& x86.c \\
\z182	& \z342	& \z227	& \z365.5	& \z117.1	& \z157.9	& x86linux.c \\[1ex]
\z\z\emph{1577}
	& \z\z\emph{2352}
		& \z\z\emph{1150}
			& \z\z\emph{2783.5}
				& \z\emph{885.0}
					& \z\z\emph{1428.1}	& \rmfamily Total\\
\noalign{\kern 2pt\hrule height 1pt}
\end{tabular}
\end{center}
\end{table}

The first column gives the time in centiseconds for compiling the
module named in the rightmost column with the monolithic version of
lcc. The second and third columns give the compilation times for
rcc and pass2. Thus, for example, the fourth row shows that
the monolithic compiler compiled \texttt{dag.c} to assembly language
in $33$~csecs., and rcc and pass2 accomplished the same task in
$61+31=92$~csecs.

The fourth column gives the size of each module's pickle in
kilobytes. By way of comparison, the fifth column gives the size of
the corresponding unoptimized object file produced by the Microsoft Visual C/C++
compiler. Pickles contain complete symbol-table information, so
perhaps a more meaningfull comparison is with the sizes of object files with
embedded symbol tables, which the sixth column shows.

Input/output time dominates the compilation times. The revised rcc is
about 1.5--2 times slower than the monolithic compiler; building the
ASDL data structures and emitting them accounts for most of this
time. As detailed in the next section, rcc essentially duplicates its
data structures as it builds the ASDL representation, which costs
both time and space.

pass2 is faster than both the monolithic
compiler and rcc because it doesn't have to read and analyze the C
source code; it simply inhales the pickle, rebuilds the compiler's
data structures, and calls the back end functions. While rcc plus
pass2 adds a factor of 2--3 to the compilation time, lcc is fast
enough that this overhead is acceptable, especially in an
experimental setting; for example, the monolithic compiler compiles
itself in 15~secs., and rcc plus pass2 takes 35~secs.

Pickle sizes run about 3 times the sizes of object files and about 2 times
the sizes of object files with embedded symbol tables. Compression
could reduce pickle sizes to that of object files; for example,
compressing all of the pickles listed in Table~\ref{table:results}
yields a 867~KB zip file. Each pickle includes the symbol-table
entries from the common header files included by each module.
Pickle sizes could also be reduced by
emitting these symbol-table data into a separate pickle and
omitting them from the per-module pickles.

\section{Evaluation}

Retrofitting lcc to use ASDL
highlighted some strengths and weaknesses in both ASDL and lcc.
One of the somewhat unexpected strengths of ASDL is that it helps
find bugs. Writing the ASDL back end revealed two related
long-standing bugs in lcc. The first is illustrated by the following
code.
\begin{quote}\singlespace
\begin{alltt}
f(void) \{ extern int x; \dots \}
int x;
\end{alltt}
\end{quote}
The two declarations for \texttt{x} refer to the same identifier.
The error is that lcc created two symbol-table entries for
\texttt{x}: One was created at the extern declaration for
\texttt{x} and used when compiling the body of \texttt{f}, and the
other one was created at the top-level declaration for \texttt{x} and
used thereafter, including announcing the definition of \texttt{x}
via the code-generation function \texttt{global} (see
Fig.~\ref{fig:functions}). It was intended that there be only one
symbol-table entry for \texttt{x}. These symbol-table entries had
identical \emph{contents}, and all of the existing back ends examined
only the contents. The ASDL back end, however, used the pointer to
the symbol-table entry as a handle to the corresponding ASDL
\texttt{symbol} type (see Fig.~\ref{fig:grammar}) and thus
erroneously created two \texttt{symbol}s. References to \texttt{x}
from within \texttt{f} referred to the wrong \texttt{symbol} and thus
the generated code was incorrect---it was if the code had been
written as
\begin{quote}\singlespace
\begin{alltt}
f(void) \{ static int x; \dots \}
int x;
\end{alltt}
\end{quote}

The second bug adds another twist to the first bug and is illustrated
by the following code.
\begin{quote}\singlespace
\begin{alltt}
static int x;
f(void) \{ extern int x; \dots \}
\end{alltt}
\end{quote}
Again, lcc erroneously created two symbol-table entries when one was
expected. It also changed the storage class of the \texttt{x}
declared within \texttt{f} to be static when it was announced
to the back end, then changed it back to \texttt{extern}. As a
result, the \texttt{x} appeared to be static when declared and extern
when used. On targets that handle statics differently than
externs, pass2 emitted incorrect code.

ASDL exposed some awkward binding times in the lcc code-generation
interface, which required revising the implementation.
lcc compiles the function
\begin{quote}\singlespace
\begin{alltt}
f(x, y) char x; int y; \{ \dots \}
\end{alltt}
\end{quote}
as
{\newcommand{\x}{$x^\prime$}\newcommand{\y}{$y^\prime$}%
\begin{quote}\singlespace
\begin{alltt}
f(? int \x, ? int \y) \{ ? char x = \x; ? int y = \y; \dots \}
\end{alltt}
\end{quote}
lcc generates two symbol-table entries for each parameter: one for
the parameter as passed by the caller---\x\ and \y\ in
the code above---and one for the parameter as seen by the
callee---\texttt{x} and \texttt{y} above. It generates assignments of
the caller parameter to the corresponding callee parameter if their
types differ or if their storage classes differ. In the example
above, in which the occurrences of \verb|?| denote storage classes,
the char parameter \texttt{x} is promoted and passed as an
int, so the types of \texttt{x} and \x\ differ. Back ends can
change the storage class of caller and callee parameters to reflect
target-dependent calling conventions. On the MIPS, for example,
\y\ is passed in a register, so an assignment to \texttt{y} is
generated if \texttt{y} lands in memory.

}

The binding time problem is that rcc doesn't have the information
necessary to determine whether or not to generate these assignments.
Storage class information is known only to the back end and thus
isn't known until pass2 runs. The solution was to move the code that
generates these assignments into pass2.

A similar problem arises in common subexpression elimination (CSE),
but requires extra work by both rcc and pass2.
lcc does CSE on extended basic blocks, but it needs to know something about 
register assignments before it hoists an rvalue into a temporary.
For example, in the expression
\begin{quote}\singlespace
\begin{alltt}
a = b*c + b*d
\end{alltt}
\end{quote}
the rvalue of \texttt{b} is a common subexpression. lcc copies
\texttt{b} to a temporary, but only if the temporary is in a register
and \texttt{b} isn't. Again, rcc does not have the data necessary to make that
decision, because the back end has the final say on storage classes.
So, rcc makes a conservative assumption and generates temporaries for
all multiply referenced rvalues, and pass2 eliminates those that don't pay.

One of the flaws in ASDL is that it can lead to a duplication of data
structures, which is perhaps should be expected when modifying an
existing compiler to use ASDL. lcc builds numerous data structures to
represent the C source program, e.g., symbol-table entries, tree
nodes, strings, etc. Most of the code in the ASDL back end is devoted
to building copies of these data structures---that is, building a
different, but logically equivalent representation for nearly
everything. All this duplication could be avoided if ASDL were used
at the outset to define all the important data structures, but this
approach would have required a much more drastic revision of lcc.

Perhaps the biggest nuisance in using ASDL is dealing with non-tree data 
structures, e.g., by using uids for symbols and types.
These are common and there should be a better way to handle them, or
at least some more built-in support for defining and referencing
them.

The lcc ASDL grammar is `ambiguous;' that is, it permits construction
of type instances that do not represent valid lcc code-generation
interface structures. For example, the grammar in
Fig.~\ref{fig:interface} permits a \texttt{Function} whose
\texttt{codelist} field includes another \texttt{Function}. This
sequence of calls never occurs in lcc. Similar comments apply to the
\texttt{CVT} and \texttt{Binary}, \texttt{Unary}, \texttt{Compare}
constructors: any operator could be given as the \texttt{op} field.
Ambiquity shortens grammars, much the same way as an ambiguous YACC
grammar is smaller than a non-ambiguous one. While few bugs could be
attributed directly to using an ambiguous grammar, the savings
probably isn't worth it. A non-ambiguous ASDL grammar, which might be
no more than 50\% longer, would catch more errors at compile-time and
it would document the semantics more accurately.

\section{Conclusions}

Revising lcc to use ASDL was, overall, straightforward, and the
resulting components---rcc and pass2---provide an improved platform
for compiler-related research using lcc. It is now possible to insert
additional passes into the compilation pipeline without modifying or
even understanding the front and back ends.
The obvious first candidate is a global optimization pass. Adding an
optimizer will surely identify weaknesses in the current 
ASDL grammar. For example, it is likely that additional data
structures, such as a flow graphs, will be needed. The optimizer could
build a flow graph itself, but it may prove useful to add these kinds
of generally useful structures to the pickles.

Fortunately, ASDL can accommodate additions gracefully. A pickle
consists of one \emph{or more} instances of ASDL types. Currently,
lcc pickles hold just an instance of \texttt{program} defined in
Fig.~\ref{fig:grammar}. Other passes can append instances of
additional types to pickles and use these instances without affecting
pass2, because pass2 reads only the instance of \texttt{program}.

ASDL is equivalent to Document Type Declarations (DTDs) in
XML~\cite{goldfarb:prescod:98}, so it is 
natural to wonder if the increasing investment in XML tools
can be leveraged to provide better compiler infrastructure tools.
As a first step,
the ASDL pickle readers and writers have been modified to emit
pickles in XML instead of in the original binary format.
These pickles are necessarily huge, because they are written in
readable ASCII, but they compress to approximately the sizes
suggested in Table~\ref{table:results}; for example, the
XML pickles for all of the modules listed in
Table~\ref{table:results} compress into a 1394~KB zip file. They can, however, be
examined and processed by generic XML browsers and editors, which
obviates the need for ASDL-specific tools.

Work is also underway on the minimal support for non-tree
data structures. XML supports \texttt{ID} and \texttt{IDREF}
`attributes;' these provide a way to name an instance of a type and
to refer to it from instances of other types, which is essentially
identical to the use of uids in the lcc ASDL grammar.
Similar features may be added to ASDL.

Finally, ASDL is an ideal way to specify abstract data types and
application programming interfaces (APIs),
independent of whether or not they are going to be pickled. ASDL
grammars are compact, language-independent, and hide implementation
details. Debugging an ASDL grammar is usually much easier than
debugging the corresponding handwritten code. With sufficient care,
ASDL grammars might help simplify both the implementations of APIs
and their tedious language-specific descriptions.

\section{Acknowledgements}

Daniel C. Wang wrote asdlGen and responded promptly when rcc and
pass2 exposed bugs. He also wrote the XML pickler mentioned in the
last section.

{\singlespace\small
\bibliography{refs,cs}

}

\begin{flushleft}
{\catcode`\$=12 \footnotesize $Id: asdl.tex,v 1.11 1998/10/12 18:39:54 drh Exp drh $}
\end{flushleft}

\end{document}